\def\year{2017}\relax
\documentclass[letterpaper]{article} 
\usepackage{aaai18}  
\usepackage{times}  
\usepackage{helvet}  
\usepackage{courier}  
\usepackage{url}  
\usepackage{graphicx}  
\usepackage{wrapfig}
\usepackage{amsmath}
\usepackage{amsfonts}
\usepackage{hyperref}

\title{Automated Cloud Provisioning on AWS using Deep Reinforcement Learning}
\author{
 Zhiguang Wang \\
 Synaptiq, LLC \\
 zgwang813@gmail.com \\
 \\
 \And
 Chul Gwon \\
 BlackSky \\
 csgwon@gmail.com \\
 \And
 Tim~Oates \\
 Synaptiq, LLC\\
 tim.oates@synaptiq.ai \\
 \And
 Adam Iezzi \\
 BlackSky \\
 adami@blacksky.com \\
}

\begin{document}
\maketitle
\begin{abstract}
As the use of cloud computing continues to rise, controlling cost
becomes increasingly important.  Yet there is evidence that 30\% - 45\%
of cloud spend is wasted \cite{rightscale}.  Existing tools for cloud
provisioning typically rely on highly trained human experts to specify
what to monitor, thresholds for triggering action, and actions.  In
this paper we explore the use of reinforcement learning (RL) to acquire
policies to balance performance and spend, allowing humans to specify
what they want as opposed to how to do it, minimizing the need for
cloud expertise.  Empirical results with tabular, deep, and dueling
double deep Q-learning with the CloudSim \cite{cloudsim}
simulator show the utility of RL and the relative merits of the
approaches.  We also demonstrate effective policy transfer learning
from an extremely simple simulator to CloudSim, with the next step
being transfer from CloudSim to an Amazon Web Services physical
environment. 

\end{abstract}

\section{Introduction}
Cloud computing has become an integral part of how businesses and other entities are run today, permeating our daily lives in ways that we take for granted.  Streaming music and videos, e-commerce, and social networks primarily utilize resources based on the cloud.  The ability to provision compute nodes, storage, and other IT resources using a pay-as-you-go model, rather than requiring significant up-front investment for infrastructure, has transformed the way that organizations operate.  The potential drawback is that groups leveraging the cloud must also effectively provision their resources to optimize the tradeoffs between their costs and the performance required by their service level agreements (SLAs).

Whereas organizations typically hire cloud experts to determine the
optimal strategies for provisioning their cloud resources, in-depth
understanding of not only cloud management, but also the business
domain are required to effectively perform this task.  Sadly, the
tools that users have to manage cloud spend are relatively meager.
Consider Amazon’s Auto Scaling service.  Users create an Auto Scaling
Group (ASG), which is a collection of instances described in a Launch
Configuration, and then choose from a rather limited set of options
for managing that infrastructure.  For example, one simple option is
to specify the desired capacity for the ASG, which
results in machines being replaced in the group when they fail a
periodic health check. 
	
A more hands-off approach is dynamic scaling in which the user defines
alarms that monitor performance metrics such as CPU utilization,
memory, or network I/O, as well as policies that make changes to the
ASG in response to alarms.  Those changes can include
adding/removing instances from the group, specified as a percentage or
a fixed number, in either one large increment/decrement or in a series
of steps, with user defined upper and lower bounds on group size
enforced.

There are two primary problems with the standard approach.  The first,
and most important, is that it requires users to specify how to
achieve their goals, without any meaningful guidance.  Note that it is
easy to specify the goal: I want to reduce spend; or, perhaps
more realistically, I want to reduce spend as much as possible without
increasing average user response time by more than 2\%.  

What's needed is an automated way to learn how to achieve
spend/performance goals specified by users.  That is, we need to let
humans do what they do best - define goals - and let machines do what
they do best - use data to discover how to achieve them.  In this
paper, we explore the use of Reinforcement Learning (RL) for cloud
provisioning, where users specify rewards based on cost and
performance to express goals, and the RL algorithm figures out how to
achieve them.   Empirical results with tabular, deep, and dueling
double deep Q-learning with the CloudSim \cite{cloudsim}
simulator show the utility of RL and the relative merits of the
approaches.  We also demonstrate effective policy transfer learning
from an extremely simple simulator to CloudSim, with the next step
being transfer from CloudSim to an actual AWS cloud
environment. 

Although there are several companies offering Infrastructure as a
Service (IaaS), we focused on AWS, given that it
currently controls the largest market share out of the possible
providers.  Our work includes an AWS environment that can be used for
reinforcement learning studies as well as initial results in deploying
our models onto AWS.  A CloudFormation script with our AWS environment
is available on GitHub at \url{https://github.com/csgwon/AWS-RL-Env}.

\section{Learning Environment Setup} 
Although there are several ways to affect cloud costs, we focused our attention on Auto Scaling, where you can automatically modify the number of compute instances to adjust to changing load on your application.  These instances can be scaled manually, on a particular schedule (increase during the work week and decrease over the weekends), or dynamically (scaling based on thresholds on the metrics available from AWS).  Our study compared dynamic scaling using thresholds against reinforcement learning algorithms using the AWS metrics as state variables.

RL is a field that sits at the intersection of machine learning and
optimal control.  An RL agent is embedded in an environment.  At each
time step $t$, the agent observes the state of the environment, $s_t$, chooses
an action to take, $a_t$, and gets a reward and a new state, $r_{t+1}$
and $s_{t+1}$,
respectively.  The goal is to choose actions to maximize the sum of
future rewards.  The life of an RL agent is a sequence of observed
states, actions, and rewards, and RL algorithms can use such sequences
to learn optimal policies.  That is, RL algorithms can learn to pick
actions based on the current state that lead to ``good'' states and
avoid ``bad'' states.  Therefore, we need to specify states, actions,
and rewards for the cloud provisioning domain.
 
\subsection{State Variables}
The initial constraint that we applied to the selection of state
variables was to restrict ourselves to Hypervisor-level metrics
provided by CloudWatch
(\url{http://docs.aws.amazon.com/AmazonCloudWatch/latest/monitoring/CW_Support_For_AWS.html}).
Whereas more detailed system information could be obtained that is  relevant to the state, our goal was to introduce minimal disturbance to an existing AWS environment.  The final observables that we used for the state consisted of the number of instances along two instance-level CloudWatch metrics (CPUUtilization, NetworkPacketsIn), and two elastic load balancer-level metrics (Latency, RequestCount).  Additional CloudWatch metrics are available, but we removed those that were highly correlated to the ones that are used for this study.  These metrics are provided by AWS at 5 minute intervals, so this is the interval of the step used for our reinforcement learning system.

\subsection{Reward}
The reward was defined by the cost of the provisioned resources and an
additional graduated penalty for high CPU utilization: 3 times the
instance cost for 70-79\% CPU utilization, 5 times for 80-89\%, and 10 times for 90\%+.  The cost of the instance was determined using our AWS cost model, discussed in the following section.  The penalties applied for high utilization will typically depend on the SLA between the service provider and the customer, and would need to be adjusted to take this into consideration.

\subsection{AWS Cost Model}
We initially investigated using AWS detailed billing to provide feedback on the cost of the utilized resources.  Although the detailed billing report breaks down the cost of each resource by hour (which would have been more coarse than our 5 min interval to begin with), the actual report is only generated a few times a day, which would make it infeasible for doing updates of the reinforcement learning system.  As a result, we created an AWS cost model that would allow reasonable estimates of costs and provide them at 5-minute intervals.

For each instance, AWS charges hourly, rounding up any partial hour, with costs varying by instance type and size.  Our model consisted of a lookup table for the various instances, and provided a front-loaded cost for the provisioned instance (the RL system would see the full price at the first step, and then no cost for the next 11 steps, based on our 5-min step interval).  Although AWS allows 20 instances by default for many of their EC2 instances (with increased provisioning available if requested), for our study we restricted the total number of allowable instances to 10.
 
\subsection{Actions} 
To establish a baseline from which to compare our reinforcement learning algorithms, we implemented a threshold-based algorithm on CPU utilization for scaling the number of instances in our Auto Scaling Group (ASG).  There are additional ways of controlling an ASG, such as thresholding on multiple metrics, or setting schedules if anticipated spikes in usage are understood, but these were not included in this study.  The action space used for our threshold-based algorithm was as follows:
\begin{itemize}
	\item Add two instances for CPUUtilization $>$ 90\%
	\item Add one instance for CPUUtilization $>$ 70\%
	\item No action for CPUUtilization between 50-70\%
	\item Subtract one instance for CPUUtilization $<$ 50\%
	\item Subtract two instances for CPUUtilization $<$ 30\%
\end{itemize}

\section{Approach} 
\subsection{Tabular Q-Learner}  
As an early baseline of our reinforcement learning approaches, we implemented a tabular-based Q-learning algorithm \cite{sutton2011reinforcement}, with the off-policy update given as follows:

\begin{equation}
\begin{split}
Q(S_t,A_t) & \leftarrow  Q(S_t,A_t) + \\
&\alpha[R_{t+1} + \gamma \max_{a} Q(S_{t+1}, a) - Q(S_t,A_t)]
\label{eq:qlearn}
\end{split}
\end{equation}
where  $\alpha \in (0,1]$ is the step-size, and $\gamma \in (0,1]$ is the discount factor.  We performed a simple grid search to optimize these parameters and used a value of 0.1 for $\alpha$ and 0.99 for $\gamma$.  The state variables were discretized using non-linear binning to reduce the number of entries in the action-value table.  Though the CPUUtilization (0-100\%) and number of instances (1-10) had natural bounds, the other variables did not.  For these, an empirically-determined final bin was used that was an order of magnitude greater than what would be expected from our tests.  This limitation could be an issue with pursuing this type of Q-learning algorithm and had effects on our results, but it was sufficient for performing initial prototyping for the simulation and AWS environments.

\subsection{Convolution on Multivariate Time Series} 
\begin{figure*}[ht]
	\centering
	\includegraphics[width=0.95\textwidth]{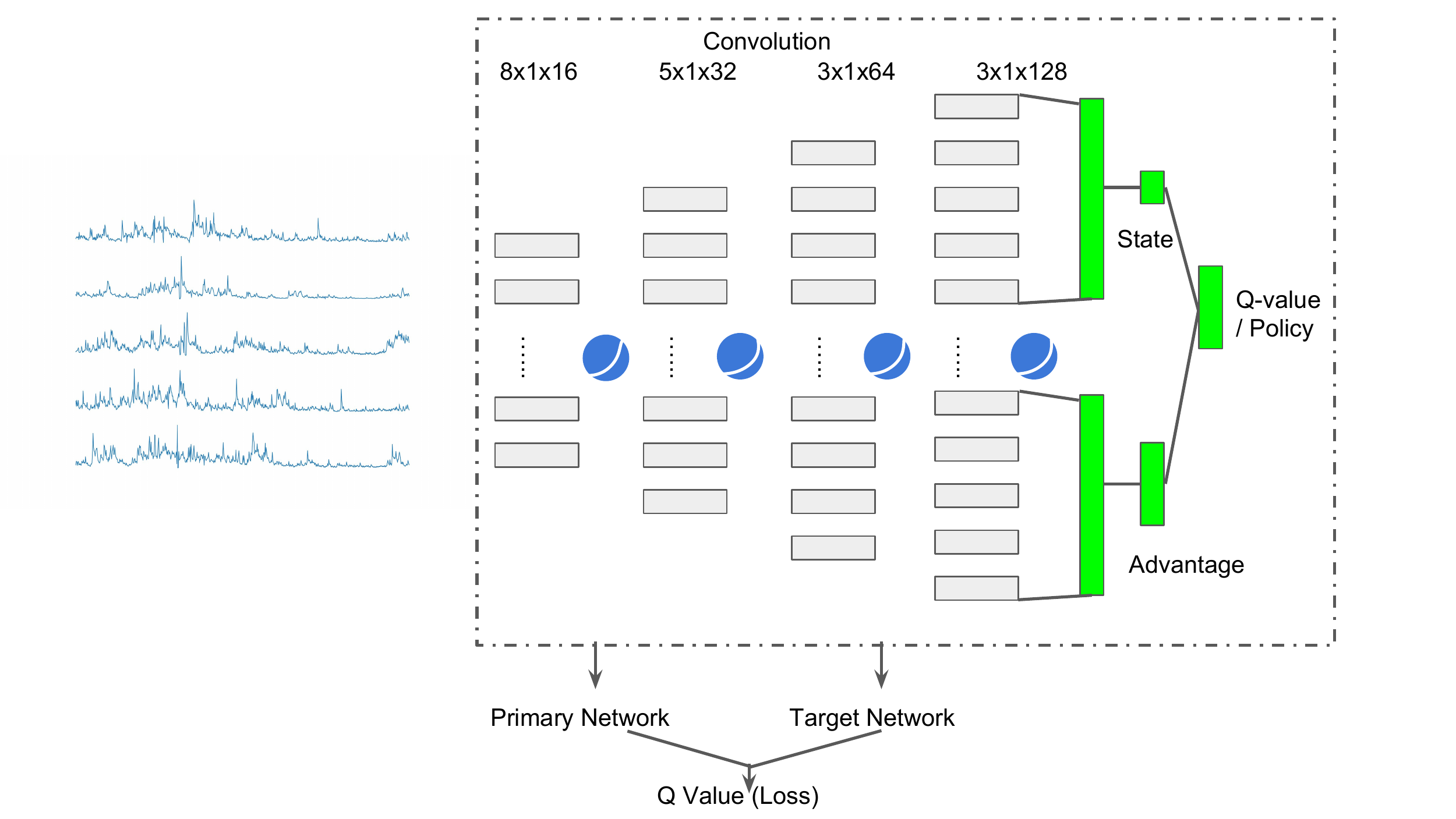}
	\caption{The network architecture of the Double Dueling Deep-Q networks.It contains four convolution layers followed by SeLU activation \protect\cite{klambauer2017self}. No pooling operations. Instead of inserting fully connected layer before final output in \protect\cite{wang2015dueling}, we halve the flattened feature maps from the last convolutional layers to compute the state and advantage respectively as we found this architecture helps to improve the stability with less parameters.}
	\label{fig:D3Q}
\end{figure*}

We also used deep convolutional neural networks as function
approximators to deal with the multivariate, real-valued data streams
flowing from CloudWatch.  In the proposed algorithm, the convolutional
networks with 1-D convolutions are built for local feature learning on
the multivariate time series data. In contrast to the usual
convolution and pooling that are both performed with square kernels,
our algorithm generates 1-D feature maps using deep tied convolutions
along each channel.

1-D convolution across the temporal axis effectively captures temporal
correlations \cite{wang2017time}. However, 2-D 
convolution across channels are meaningless in this case as the local neighbors of
different channels may or may not have interactions and correlations
through the time window. The channel ordering caught by 2-D
convolutions is also hard to interpret. By using the tied 1-D
convolution kernels across different channels, we implicitly force the
network to communicate among the different channels. Compared with
separating the kernel in each channel and then concatenating the
flattened features, using tied kernels enables equivalent or better
performance with significantly fewer parameters. The network structures
are shown in Figure \ref{fig:D3Q}.

\subsection{Deep Q-Network (DQN)}  

Though tabular Q-learning is easy to implement, it doesn't scale as the number of
possible states grows large or is infinite, as with cloud
provisioning where the states are continuous.  We instead need a way
to take a description of our state, and produce Q-values for actions
without a table. A deep Q-Network (DQN) \cite{mnih2015human} can be
used as a function approximator, taking any number of possible states
that can be represented as a vector and learning to map them to Q-values
directly.  

We used a four-layer convolutional network that takes the
continuous states within a sliding window of size $K$ and produces a
vector of 5 Q-values, one for each action (+2, +1, hold, -1, -2). The
Q-learning update uses the following loss function: 

\begin{equation}
\begin{split}
L_t(\theta_t) & =  \mathbb{E}_{(S, A, R, S')} 
[(R_{t+1} +  \\
&\gamma \max_{a} Q(S'_{t+1}, A';\theta^-_t) - Q(S_t,A_t;\theta_t))^2]
\label{eq:DQN}
\end{split}
\end{equation}

$\theta$ is the parameters of the Q-network at $t$, and $\theta^-$ is
the network parameters used to compute the target at $t$. $S'$ is the state obtained after choosing the optimal action $A_t$.
	
The key difference is that instead of directly updating a table, with a neural network we will be able to predict the Q-value directly. The loss function is the difference between the current predicted Q-values and the ``target'' value. Consequently, Q-target for the chosen action is the equivalent to the Q-value computed in Equation \ref{eq:qlearn}.

\subsection{Double Dueling Deep Q-Network (D3QN)}  

In Q-learning and the original DQN, the max operator uses the same
values to both select the policy and evaluate an action, which can
lead to overly optimistic value estimates. That is, at every
step of training, the Q-network's values shift, and if we are using a
constantly shifting set of values to adjust the network values, then
the value estimations can easily spiral out of control. Especially for
control systems on time series data with a sliding window, we found
that using a single Q-network can lead to very unstable performance
during training as it fell into a feedback loop between the target
and estimated Q-values.  

One possible solution is the utilization of a second network during
the training procedure \cite{van2016deep}, i.e., a Double
Q-Network. This second network is used to generate the target-Q values
that will be used to compute the loss for every action during
training. Indeed, the target network's weights are fixed, and only
periodically updated to the primary Q-network's values to make training
more stable. In our experiments, instead of updating the target
network periodically and all at once, we updated it
frequently but slowly \cite{lillicrap2015continuous}, as we found that
doing so stabilized the training process on multivariate time series. 

The Q-values $Q(S,A)$ that we were discussing correspond to how
good it is to take a certain action given a certain state. This action
given the state can actually be decomposed into two more fundamental
notions of assets - a value function $V(S;\theta)$, which says how
optimal it is to be in any given state, and an advantage function
$A(S,A;\theta)$, which shows how much better taking a certain action
would be compared to the others. Thus, the Q-value can be simply
decomposed as:

\begin{equation}
Q(S,A;\theta) =  V(S;\theta) + A(S,A;\theta)
\label{eq:Dueling1}
\end{equation}

In real reinforcement learning settings, the agent may not need to
care about both the value and advantage at any given time. In other 
words, for many states, it is unnecessary to estimate the value of each
action choice. For example, in our AWS control setting, knowing
whether to add or reduce the number of instances only matters when a
collision among CPU, hard disk or even bandwidth is eminent. In some
states, it is of paramount importance to know which action to take,
but in many other states the results are not coupled with the action
choices and it doesn't really make sense to think of the value of
taking the specific actions being conditioned on anything beyond the
environmental (AWS) state within. The single-stream architecture also
only updates the value for one of the actions, where the values for all
other 
actions remain untouched. 

Given these observations, we used a Dueling
Q-network \cite{wang2015dueling} to 
achieve more robust estimates of state value by decoupling it from the
necessity of being attached to specific actions, using
Equation \ref{eq:Dueling1}. But the $Q$ value cannot be recovered from
$V$ and $A$ uniquely. This lack of identifiability leads to poor
practical performance when this equation is used directly. We can
force the advantage function estimator to have zero advantage at the
chosen action by implementing the forward mapping:

\begin{equation}
\begin{split}
Q(S,A;\theta) =  
&V(S;\theta) + \\
&(A(S,A;\theta) - max_{a \in A}A(S,A';\theta))
\label{eq:Dueling2}
\end{split}
\end{equation}

The goal of the Dueling DQN is to have a network that separately computes
the advantage and value functions, and combines them back into a
single $Q$ function only at the final layer. We also removed the
two-stream fully connected layer in the original Dueling
Q-network. Instead, we can learn faster and achieve more robust
learning by synthesizing advantages and state-values directly from
the flattened feature maps from the last convolutional blocks
(Figure \ref{fig:D3Q}). 

\section{Experimental Setup}  
Given the cost and time that is associated with running on a true cloud environment, previous efforts have primarily used simulators to perform their studies and evaluate performance.  Whereas the ability to simulate is essential for efficiently evaluating the expected performance, our approach also involves running on an actual AWS environment in order to obtain a true evaluation of performance.  To this end, we created two simulation environments for performing initial prototyping, debugging, and testing as well an AWS environment.  

The API for the environments was inspired by that used by the OpenAI Gym (\url{https://gym.openai.com}), using Python as the primary programming language for ease of integration into the deep learning systems. A common API allowed us to easily change the environment used by the RL system between simulators and AWS, as well as the ability to take architectures used with OpenAI Gym for our study.

We simulated network traffic to all environments, simulated and AWS,
using counts obtained from the LBL-CONN-7 dataset (see the Simulating Network Traffic section).

\subsection{Simulation Environments}  
Model testing and initial evaluation were performed using a simulated environment, which was particularly important given that our step interval was 5 minutes.  We used two different simulation frameworks for performing our tests: a simplified Python environment for running very fast tests, and a more realistic environment based on CloudSim [1], which is a Java-based package that simulates Infrastructure as a Service (IaaS).  For our CloudSim environment, we created a wrapper in Python in order to have it conform with the API of our environments.

For CPUUtilization, NetworkPacketsIn, and RequestCount metrics, we derived models based on tests that were run on our actual AWS environment.  Particularly for CPUUtilization, this model translated into a scaling law for our simple simulator, while using the model to convert requests to Instructions Per Second (IPS), which is the input that CloudSim uses to define jobs.

Although we defined the necessary parameters in CloudSim for General Purpose (M), Compute Optimized (C), and Memory Optimized (R) AWS instances (\url{https://aws.amazon.com/ec2/instance-types/}), our initial study only used m4.large instances.  We used this instance type for our simple simulator and AWS environment as well.  Whereas the reinforcement learning techniques could also be applied to optimizing instance types and sizes, this is beyond the scope of this study.

\subsection{AWS Environment}  
\begin{figure}[ht]
	\centering
	\includegraphics[width=0.45\textwidth]{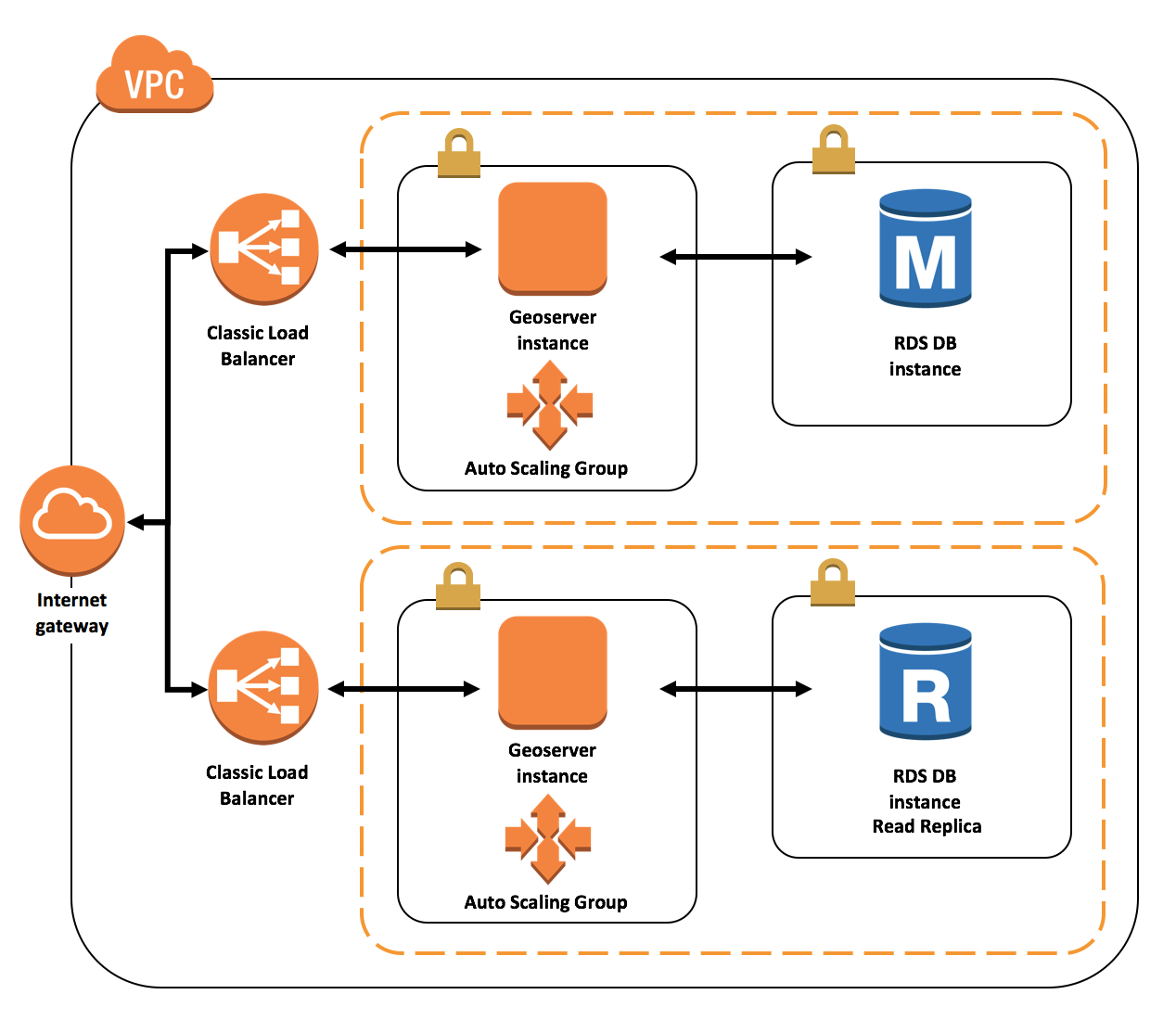}
	\caption{AWS environment.  Classic load balancer in front of an ASG of Geoserver instances.  PostGIS RDS instances with Natural Earth data.  Parallel environments using RDS read replicas with dedicated load balancers and Geoserver instances allowed concurrent tests.}
	\label{fig:aws_env}
\end{figure}

Although significant flexibility and efficiency are obtained using
simulated environments, we were particularly interested to see the
effects on an actual AWS environment.  Our architecture is shown in
Figure \ref{fig:aws_env}.  We ingested data from Natural Earth
(\url{http://www.naturalearthdata.com}) into Amazon Relational Database Service (RDS) instances using PostGIS.  We exposed the data via Geoserver (\url{http://geoserver.org}), for which custom Amazon Machine Images (AMI) were created to expose Geoserver instances in an ASG.  A Classic Elastic Load Balancer (ELB) received requests for Web Feature Service (WFS) data and forwarded them to the Geoserver instances.  Multiple parallel environments were created using RDS read replicas, allowing for concurrent testing and running the threshold-based and reinforcement learning implementations.

For our AWS environment, we used the AWS Python SDK Boto3 (\url{https://aws.amazon.com/sdk-for-python}) to poll for CloudWatch metrics and set the desired capacity on the ASG based on the action from the RL system.

\subsection{Simulating Network Traffic}\label{sec:net_traffic}  

In order to simulate calls to our web services, we used the LBL-CONN-7
data set (\url{http://ita.ee.lbl.gov/html/contrib/LBL-CONN-7.html}),
which is a 30-day trace of TCP connections between Lawrence Berkeley
Laboratory and the rest of the world.  The number of calls were binned
into 5-min intervals, scaled up by a factor of 20, and then used to
drive the RL system.  We added additional randomness by sampling from
a Gaussian distribution centered around the number of counts, with a
spread proportional to the square root of the number of counts.
\begin{figure}[ht]
	\centering
	\includegraphics[width=0.35\textwidth]{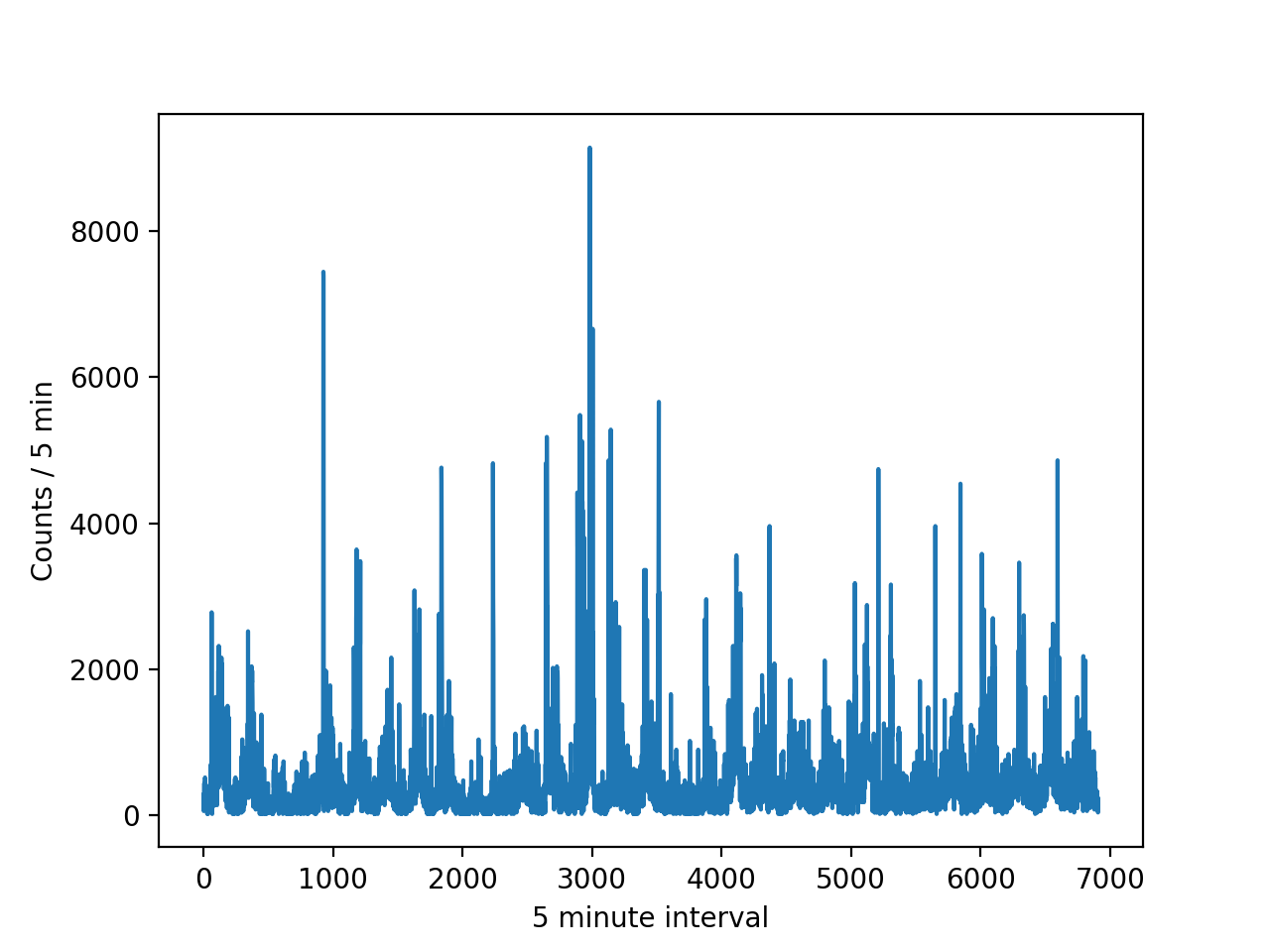}
	\caption{Number of packet traces per 5 minute interval in the LBL-CONN-7 dataset.}
	\label{fig:lbl-conn-7}
\end{figure}

In our AWS environment, we created a simple Python-based client that would use the counts to make that number of requests to our ELB for WFS data.  The simulated environment correspondingly used these counts to determine the number of calls to emulate in our system.  The size of the retrieved data varied from 200 to 5000 features.

\section{Results and Analysis}  
We present results of runs performed in our simulated environments as well as on our AWS environment.  The coarser binning used for the simulations reflects our ability to perform longer runs to see less noisy trends.  For the finer binning on the AWS environment results, we include corresponding binning from the simulation for comparison.

\subsection{Results from Simulation} 

\begin{figure}[ht]
	\centering
	\includegraphics[width=0.4\textwidth]{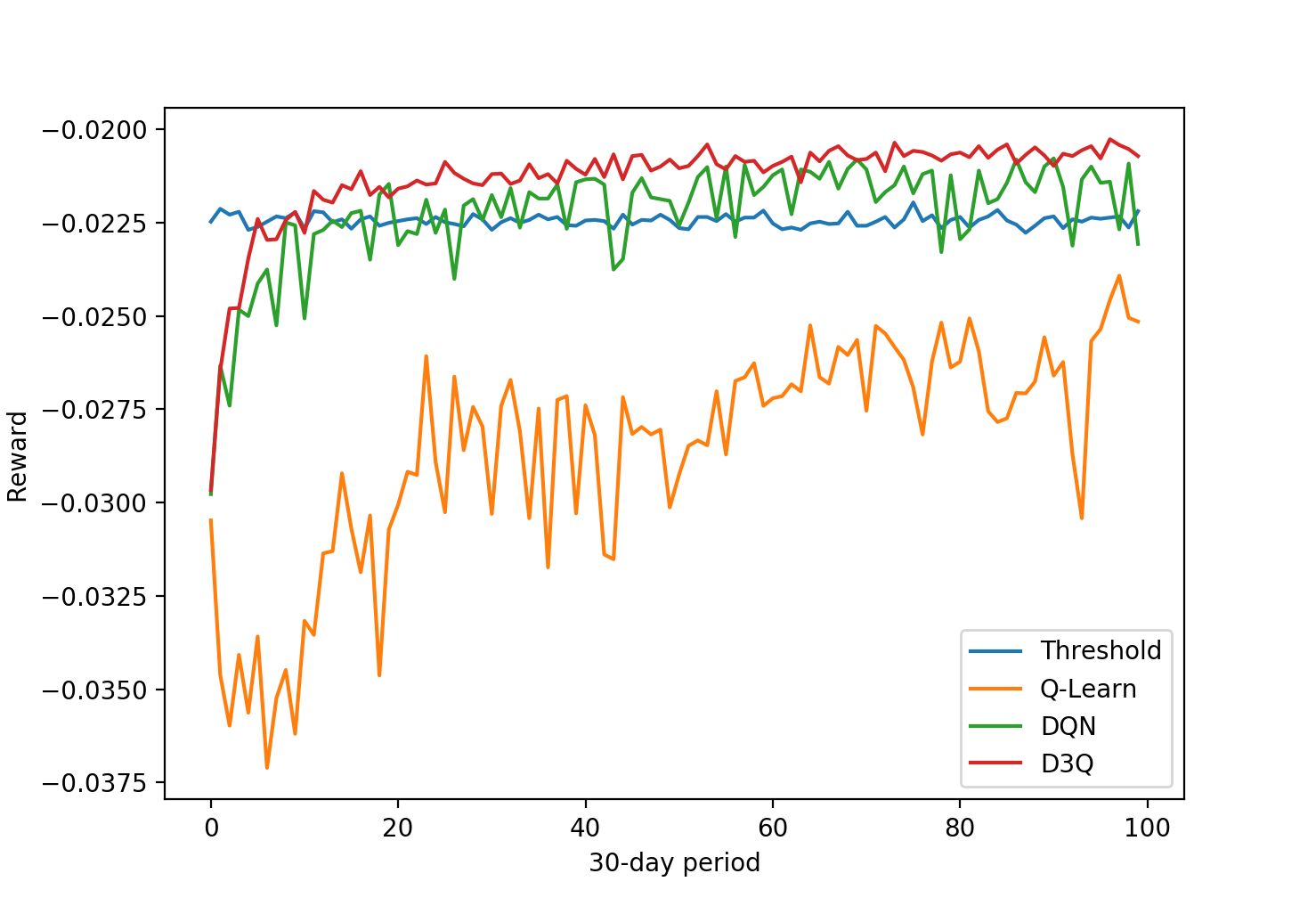}
	\caption{Comparison of the threshold-based algorithm and the reinforcement learning algorithms using simulations.}
	\label{fig:sim_comparison}
\end{figure}

Our results from simulating the threshold baseline (i.e., a
human-defined policy) and the three
reinforcement learning algorithms are shown in
Figure \ref{fig:sim_comparison}.  Each data point represents a pass
through the entire 30-day LBL-CONN-7 data set.  The performance of the
threshold baseline is essentially constant, as improvements would
require manual intervention.  Over time, the performance of the DQN
and D3QN methods surpass that of the threshold-based solution.  The
performance of the tabular Q-learner (the curve labeled Q-learn in the
figure) is poor, 
which can be attributed to many factors, including the
choice of discretization of the state variables and little
experimentation with other hyperparameters.

\begin{figure}[ht]
	\centering
	\includegraphics[width=0.4\textwidth]{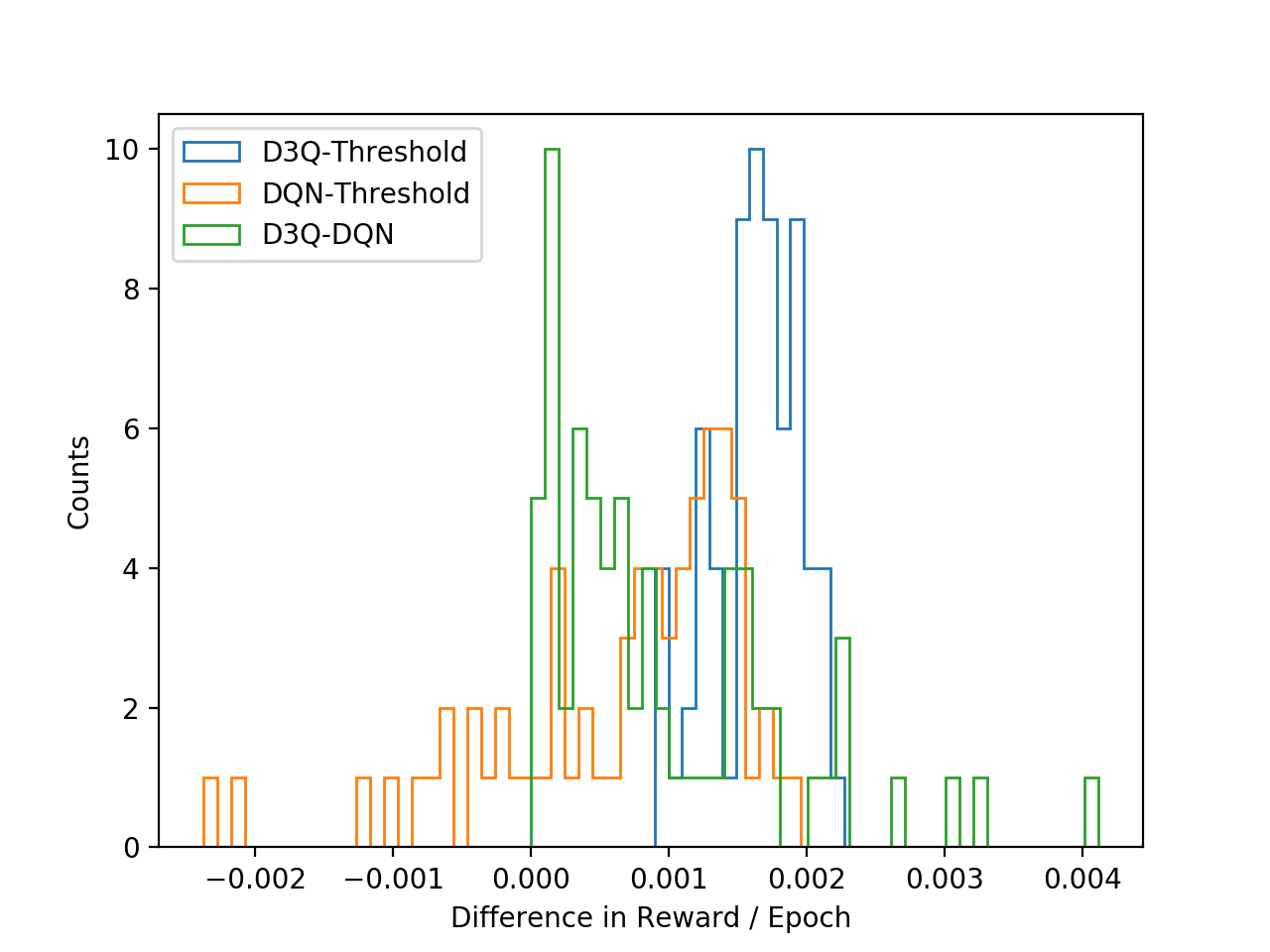}
	\caption{Difference in the average value of the reward per epoch between the different algorithms.}
	\label{fig:sim_diff}
\end{figure}

In Figure \ref{fig:sim_diff}, we also show the distribution of the
differences between the average rewards in
Figure \ref{fig:sim_comparison} for D3QN, DQN, and the threshold-based
solution.  To confirm differences in the mean values, we also ran
scipy.stats.ttest\_rel and obtained the following p-values, showing
that D3Q is statistically significantly better than the other two
algorithms, and DQN is statistically significantly better than the
threshold-based solution. 
\begin{itemize}
	\item D3Q-Threshold$=$ 4.9e-51
	\item DQN-Threshold $=$ 3.3e-8
	\item D3Q-DQN $=$ 1.1e-13.
\end{itemize}

\begin{figure}[ht]
	\centering
	\includegraphics[width=0.4\textwidth]{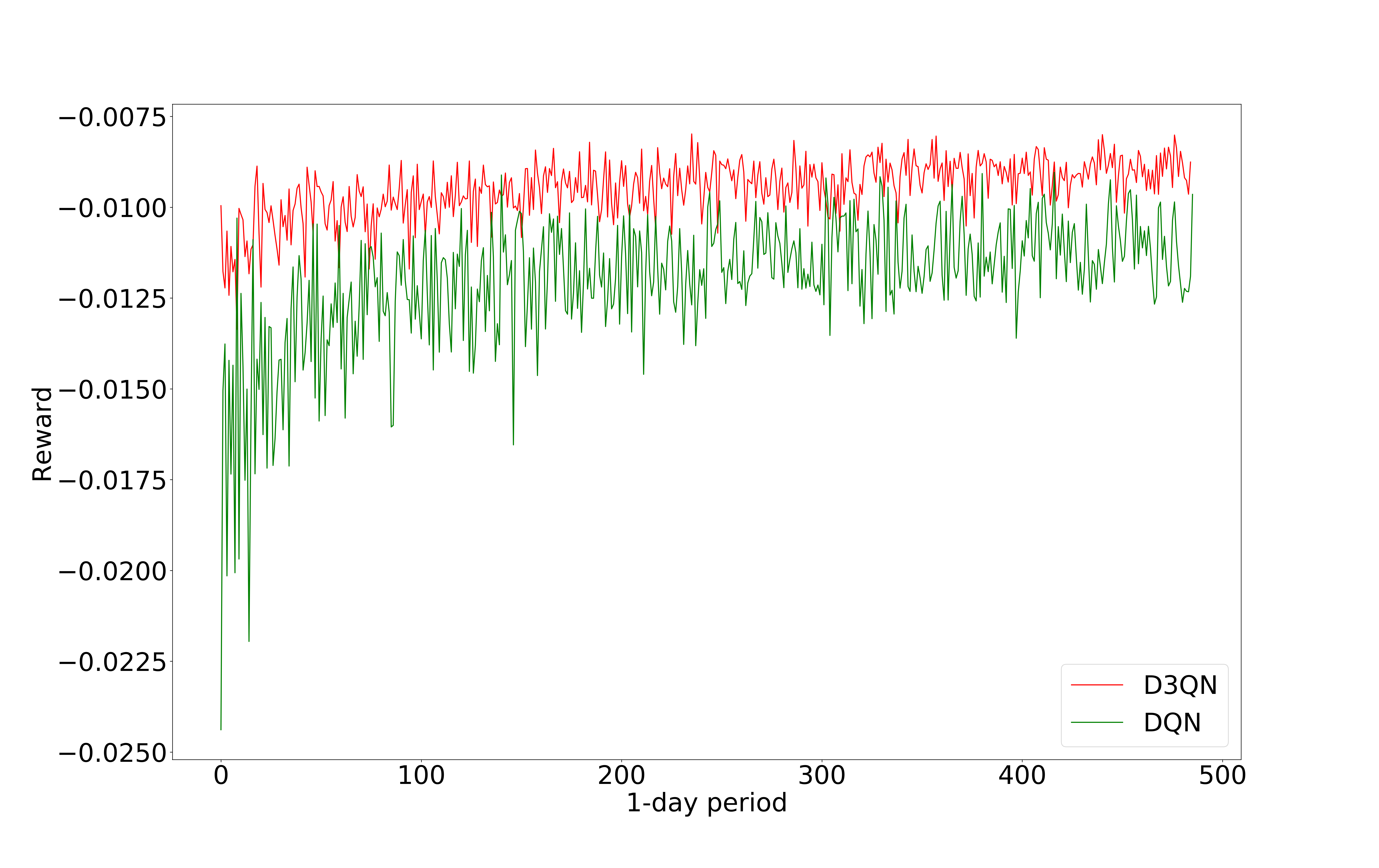}
	\caption{Training curve in the simulation environment.}
	\label{fig:Sim_plot}
\end{figure}

We compared the learning curve in the simulation environment using both DQN
and D3QN (Figure \ref{fig:Sim_plot}). D3QN dominates DQN, learns
faster, and is much more stable by decoupling the actions, states, and advantages.  

\begin{figure}[ht]
	\centering
	\includegraphics[width=0.40\textwidth]{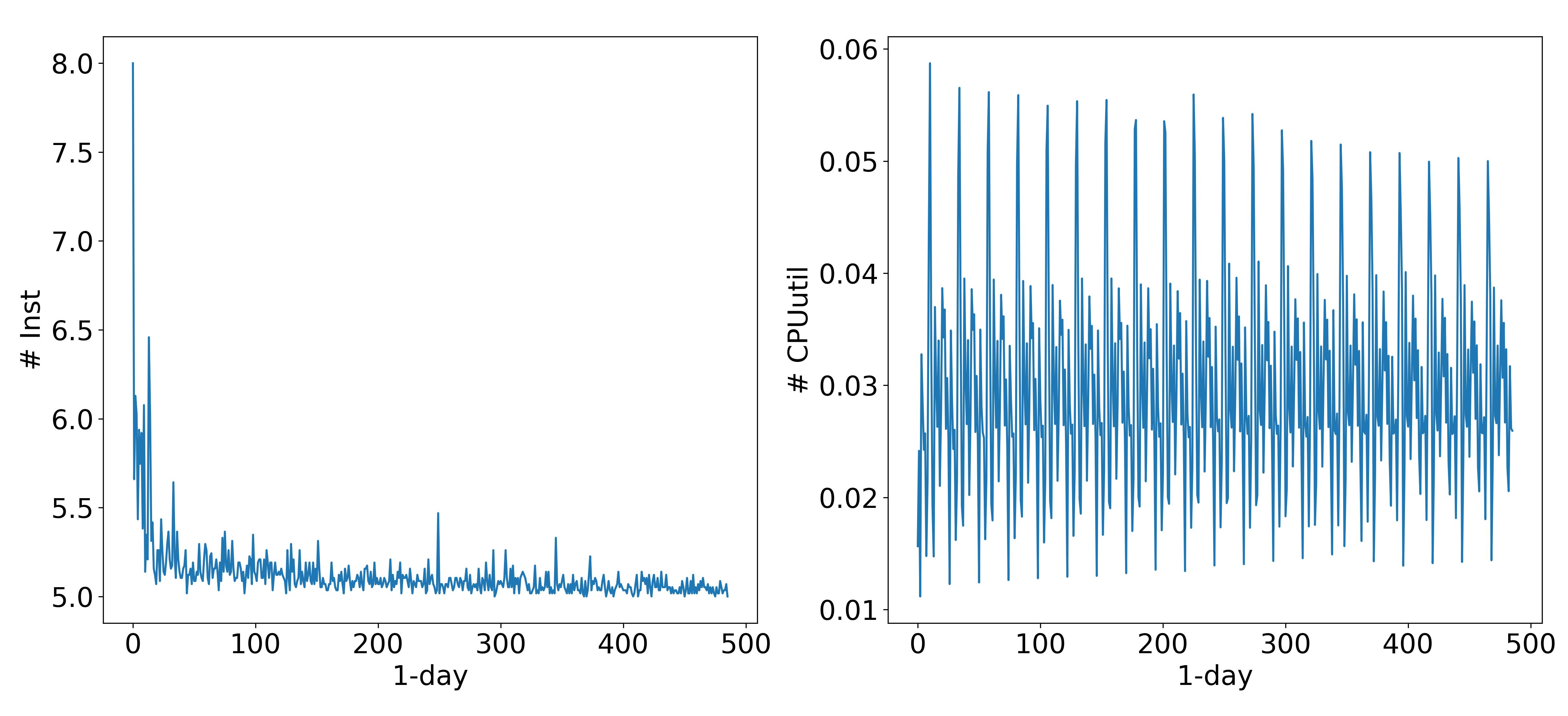}
	\includegraphics[width=0.40\textwidth]{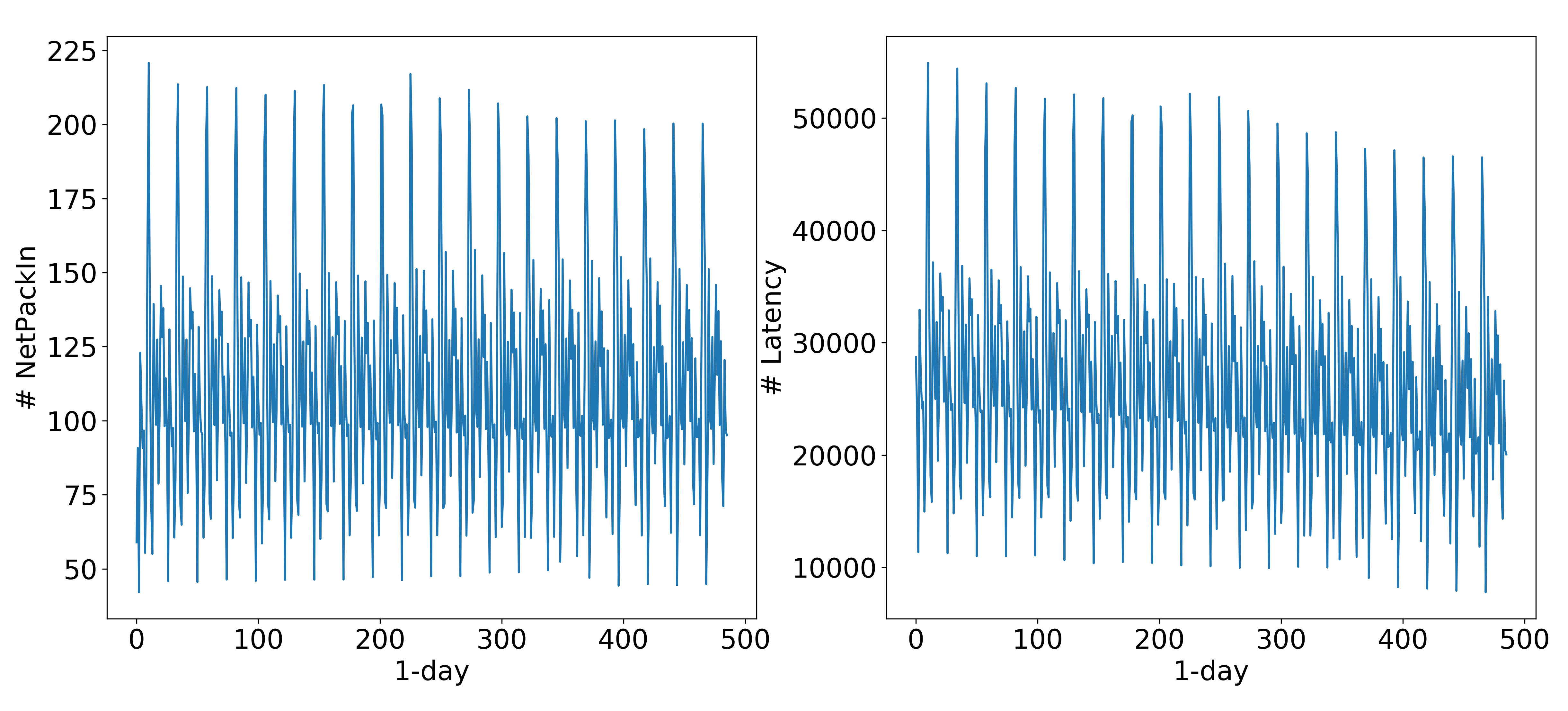}
	\caption{Learning curves of the states (InstanceNum, CPUUtilization, NetworkPacketsIn and Latency) in D3QN.}
	\label{fig:metrics_curve}
\end{figure}

We further looked at the state variables - InstanceNum,
CPUUtilization, NetworkPacketsIn, and Latency - throughout training
for D3QN (Figure \ref{fig:metrics_curve}). The number of running
instances reduces consistently during training (leftmost plot), as the policy
learned to reduce cost. The CPU utilization converges to more stable and efficient percentages (25\%-40\%),
as high CPU  utilization causes latencies and low CPU utilization
leads to more idle time and increased cost. The latency also
drops slowly despite the fact that fewer instances are running. The
D3QN learns a policy that trades off cost and efficiency.  Finally,
note that the network packets in does not
change much, as the policy has little ability to control that input,
other than reducing latency to avoid repeated requests.

\subsection{Results from AWS} 

\begin{figure}[ht]
	\centering
	\includegraphics[width=0.40\textwidth]{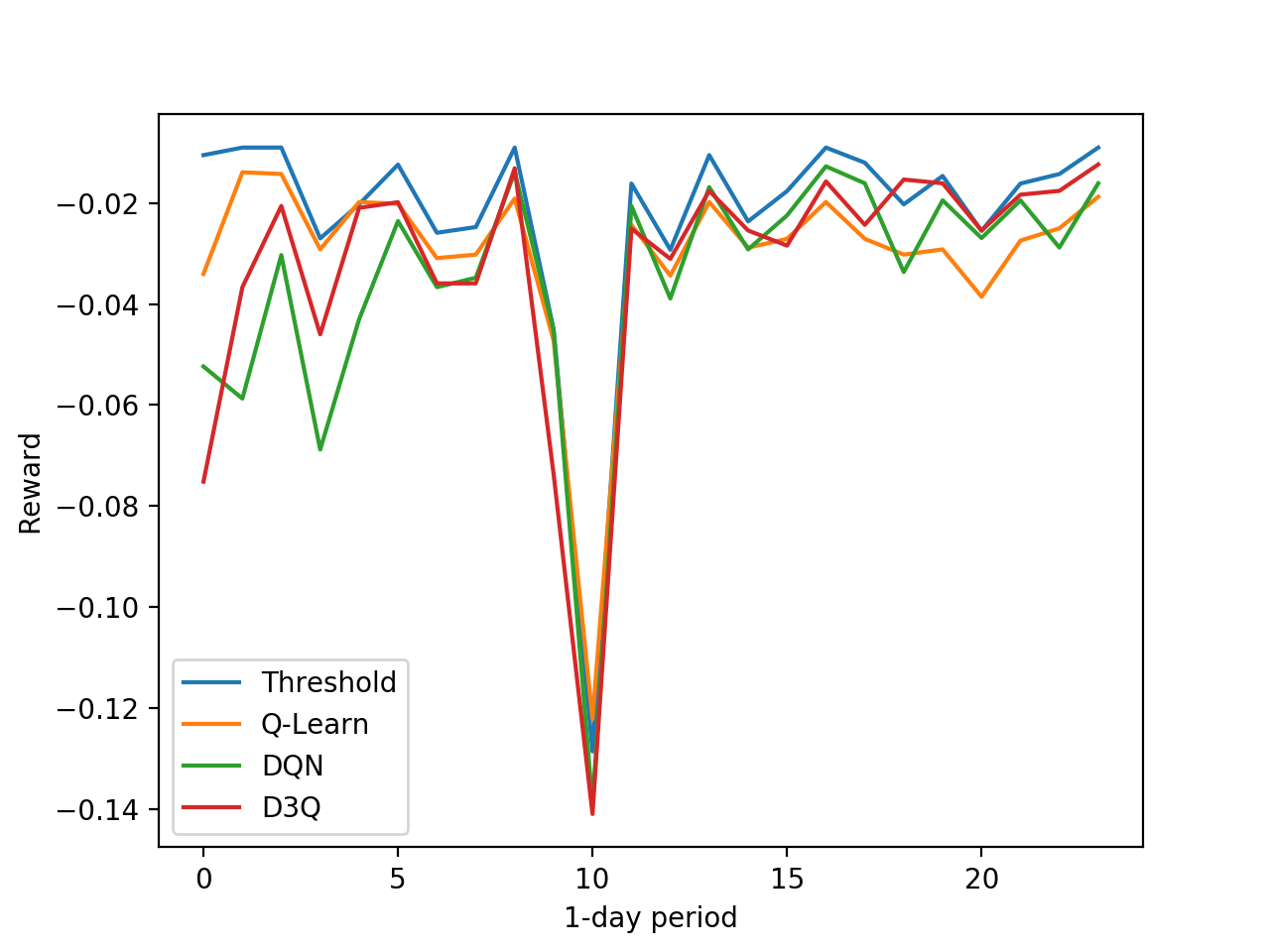}
	\includegraphics[width=0.40\textwidth]{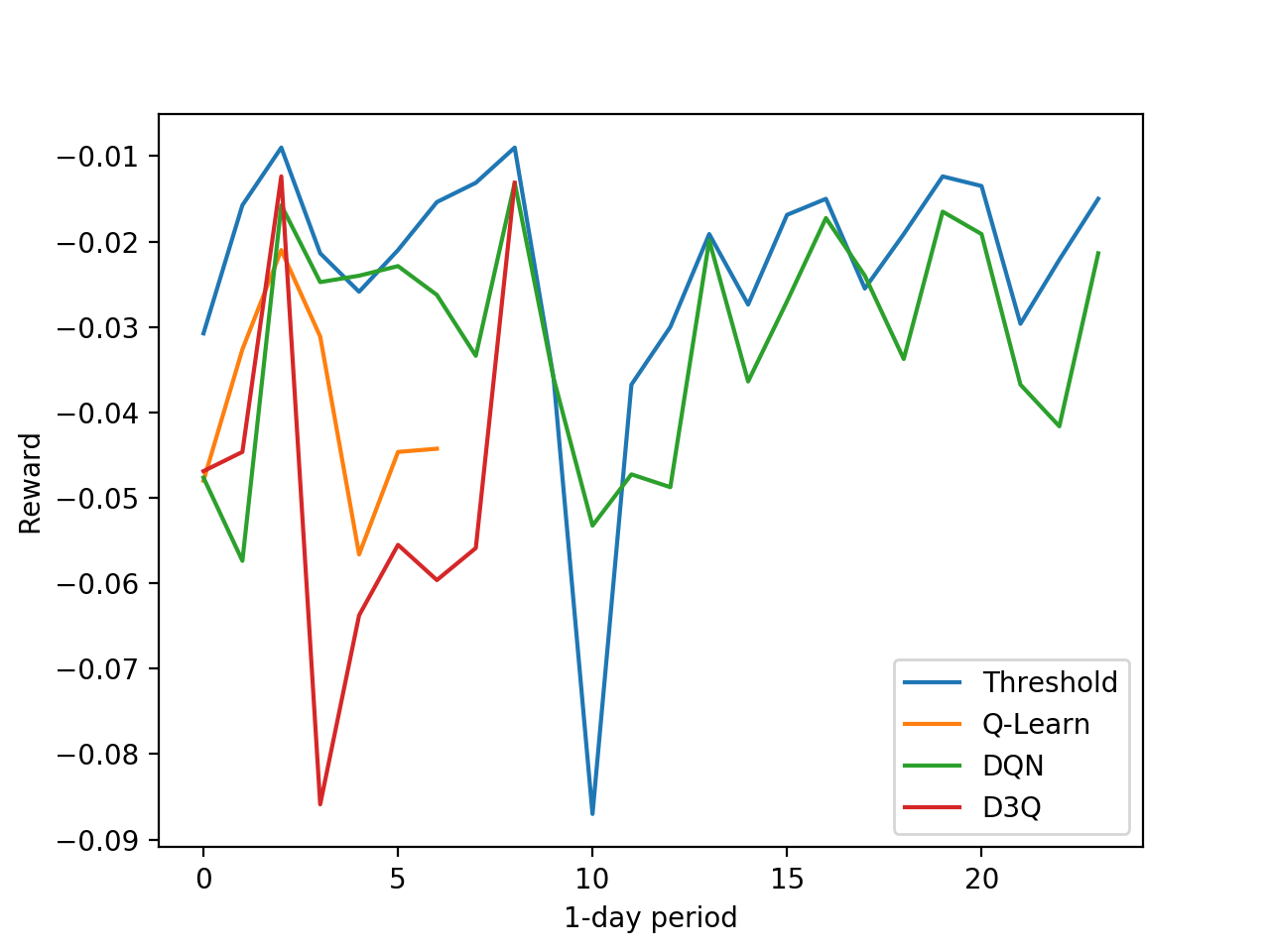}
	\caption{Comparison of the threshold-based algorithm and the reinforcement learning algorithms on simulation using 1-day binning (top) on an actual AWS environment (bottom).  Dip at Day 10 corresponds to the increase in calls from using the LBL-CONN-7 dataset (see Figure \ref{fig:lbl-conn-7}, where Day 10 are the 5 min intervals between 2880-3168) and the penalty from increased CPU utilization.}
	\label{fig:aws_results}
\end{figure}

Our results from running on AWS were limited by the amount of time we
could run.  For the threshold baseline and the DQN model, we were able
to obtain over 3 weeks of running time on the AWS system.  For the
D3QN, we obtained a week.  The tabular-based Q-learner was
intentionally run for a very short time, primarily to help initial testing
and development of the AWS environment.  Figure \ref{fig:aws_results}
shows the results of our runs, and includes the simulation runs using
the same binning to allow comparisons with expected results.  As we
observed from our simulation runs, we begin to see improvements after
the learners have been able to run on the system for longer than we
were able to for our runs.  The addition of non-zero latency values in
the actual AWS environment also affects the results.  While it is
difficult to see increasing rewards, there is clearly less variation
in reward through time as learning progresses, meaning that
performance is more stable.

\subsection{Transfer Learning}  
\begin{figure}[ht]
	\centering
	\includegraphics[width=0.45\textwidth]{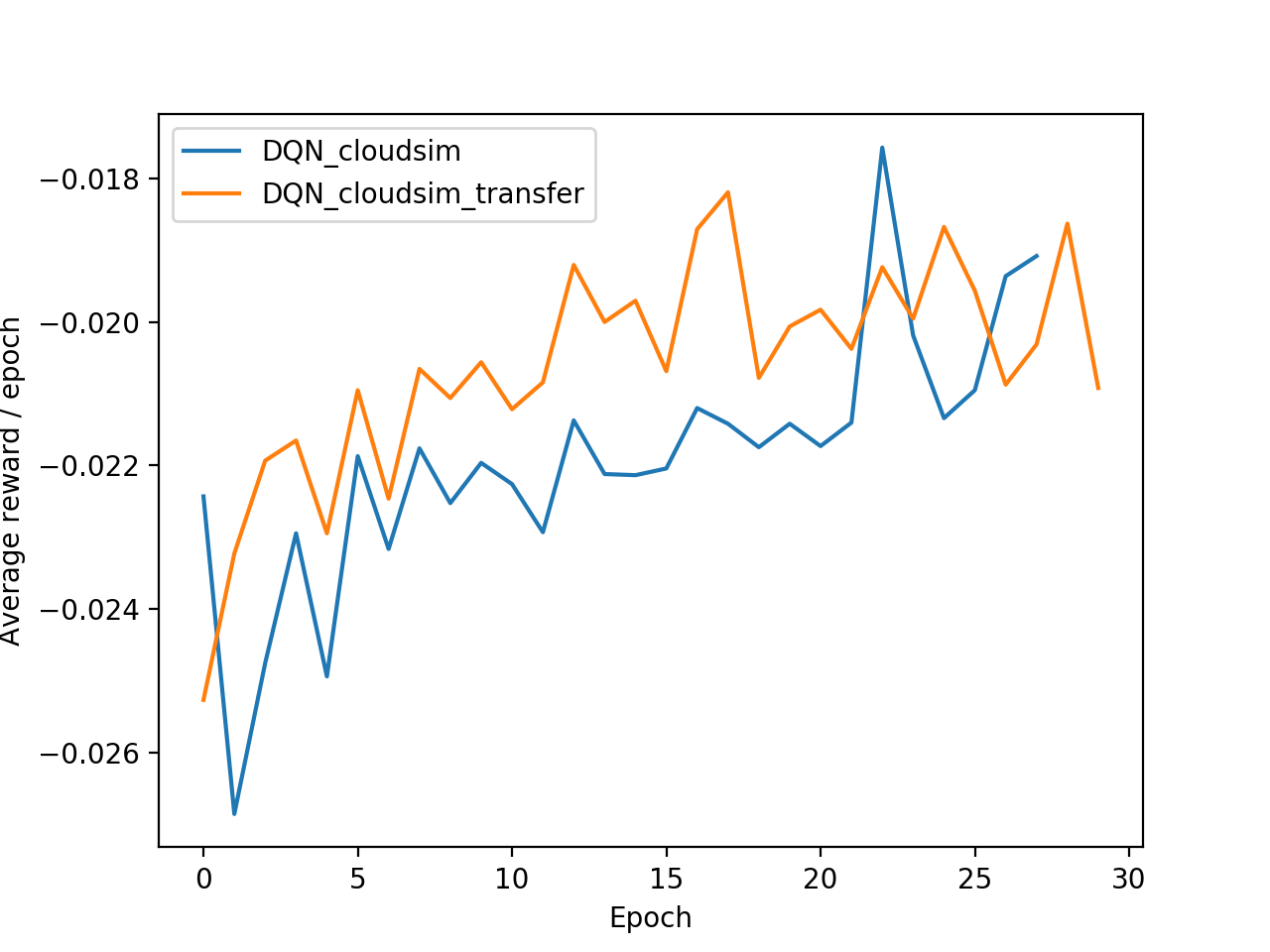}
	\caption{Results from using weights from the fast simulator to initialize the CloudSim runs.  Average reward per epoch displayed for DQN with default initialization (blue), and using fast simulation weights (orange).}
	\label{fig:transfer}
\end{figure}

An important aspect of our approach was determining whether weights
derived from simulation could be applied to the AWS environment.  The
primary relevance for this approach is to generate a policy using the
significantly more efficient simulation tools, and then allowing this
policy to fine tune on a real application.  To test this, we used
weights for our DQN generated by the fast simulation and applied them
to our CloudSim environment.  The results in Figure \ref{fig:transfer}
show the average reward versus an entire pass through the data, where
the data is defined in the Simulating Network Traffic section.
Although the average reward does seem better with the transfered
weights, with faster initial learning, further studies would be
required to determine the true feasibility of using this approach. 

\section{Related Work}  

There is limited prior work that has applied tabular-based
reinforcement learning to cloud
provisioning \cite{dutreilh2011}, \cite{mera-gomez2017}, \cite{barrett2013}, \cite{habib2016},
all of which consider very simple (discretized) states and cannot 
handle
environments consisting of large and continuous state spaces. Others
have used a feed forward neural
network \cite{xu2011,mao2016resource}, but these studies
predominantly used a simplified state space and have not explored deep
reinforcement learning methods on continuous time series.  We have
been unable to find studies that also attempted to run directly in an
AWS environment.  

Motivated by the popularity of the continuous control frameworks with
deep reinforcement learning, DQN and its variants has been studied
widely recently in Atari games, physics-based simulators and
robotics \cite{mnih2015human,levine2016end,schulman2015trust}.  We are
not aware of any work that applies the deep reinforcement learning for
optimal control on continuous time series. We are also the first to
release the standard study environments for automated cloud
provisioning.  

\section*{Discussion and Future Work}

In this paper we explore the application of reinforcement learning to
the problem of provisioning cloud resources.  Experimental results
showed that deep RL outperforms hand-crafted policies created using
existing methodologies employed by human experts for this task.
Further, Double Dueling Deep Q-learning significantly outperforms
vanilla Deep-Q policy learning both in terms of accumulated reward and
stability, which is an important quality of cloud services.

In the future we intend to do longer runs on AWS to establish real
utility and to further explore transfer learning in an attempt to
reduce the burn in time for policies where real money and customer
satisfaction are on the line.

\section*{Acknowledgments}
The authors would like to thank Andy Spohn (\url{https://andyspohn.com}) for his invaluable discussions regarding the AWS cost model and architectures.

\bibliographystyle{aaai}
\bibliography{cloudopt}

\end{document}